\documentstyle[12pt]{article}% 11pt, article report letter
\begin{document}
\vspace*{-1in}
\renewcommand{\thefootnote}{\fnsymbol{footnote}}
\begin{flushright}
TIFR/TH/96-07\\
%hep-ph/9602329\\
\end{flushright}
\vskip 65pt
\begin{center}
{\Large \bf \boldmath Charmonium production at the LHC} \\
\vspace{8mm}
{\bf K. Sridhar\footnote{sridhar@theory.tifr.res.in}}\\
\vspace{10pt}
{\it Theory Group, Tata Institute of Fundamental Research, \\ 
Homi Bhabha Road, Bombay 400 005, India.}

\vspace{80pt}
{\bf ABSTRACT}
\end{center}
\vspace{12pt}

The analyses of large transverse momentum charmonium production
at the Tevatron have shown that fragmentation of gluons is an
important production mechanism. We study large-$p_T$ charmonium
production in $pp$ collisions at the LHC, and find that due to
the copious gluon production at this energy, the gluon fragmentation
contribution completely overwhelms the fusion contribution and
the charm quark fragmentation contribution. Our analysis shows
that for $J/\psi$ production at the LHC, there is a significant
event rate even for $p_T \sim$~100~GeV. The measurement of the
cross-section at such large values of $p_T$ will provide a very
important test of the fragmentation mechanism.
\vspace{98pt}
\noindent
\begin{flushleft}
February 1996\\
\end{flushleft}

\vskip 10pt

\setcounter{footnote}{0}
\renewcommand{\thefootnote}{\arabic{footnote}}

\vfill
\clearpage
\setcounter{page}{1}
\pagestyle{plain}
The production of quarkonia has conventionally been described in
terms of the colour-singlet model \cite{berjon, br}. In this model,
a non-relativistic approximation is used to describe the binding
of the heavy-quark pair produced via parton-fusion processes, into
a quarkonium state. The heavy-quark pair is projected onto a 
physical quarkonium state using a colour-singlet projection and 
an appropriate spin-projection. It is also necessary to account 
for $J/\psi$'s produced $via$ electromagnetic decays of the 
$P$-states, $viz.$, the $\chi$'s, since inclusive $J/\psi$ production 
is measured in experiments. In the model, both the direct 
$S$-state production and the indirect production $via$ $P$-state 
decays are accounted for. This model has been successfully applied \cite{br} 
to describe large-$p_T$ $J/\psi$ production at the relatively low 
energies in the ISR experiment. In going from the ISR energy 
($\sqrt{s}=63$~GeV) to the UA1 energy ($\sqrt{s}=630$~GeV), it was
found \cite{gms} that the production of $b$-quarks and their subsequent
decay becomes an important source of $J/\psi$
production. While at the UA1 experiment it was not possible to
separate the $b$-quark contribution, the use of the silicon-vertex
detector at the CDF experiment in studying $J/\psi$ production
in $\bar p p$ collisions at the Tevatron ($\sqrt{s}=1.8$~TeV)
allows the subtraction of the $b$-quark contribution from the
total yield, thereby providing a measurement of direct
inclusive $J/\psi$ production at large-$p_T$.
The inclusive $J/\psi$ production cross-section measured by the
CDF experiment \cite{cdf} turned out to be an order of magnitude
larger than the prediction of the colour-singlet model. 

\vskip10pt
In fact, even before the experimental results from the CDF 
collaboration were available, it was shown
by Braaten and Yuan \cite{bryu} that in addition to the parton
fusion contributions taken into account in the colour-singlet
model, fragmentation of gluons and charm quarks could be an
important source of large-$p_T$ $J/\psi$ production. In the
fragmentation process, one considers the production of charmonia 
from final-state gluons or charm quarks which have large $p_T$, 
but almost zero virtuality. The fragmentation contribution is 
computed by factorising the cross-section for the process 
$AB \rightarrow (J/\psi,\chi_i) X$ (where $A,\ B$ denote hadrons)
into a part containing the hard-scattering cross-section for producing a
gluon or a charm quark and a part which specifies the fragmentation of
the gluon or the charm quark into the required charmonium state, i.e.
\begin{equation}
d\sigma (AB \rightarrow (J/\psi,\chi_i) X)
 = \sum_c \int_0^1 dz \hskip4pt
d\sigma (AB \rightarrow c X) D_{c \rightarrow (J/\psi,\chi_i)}
(z,\mu ) ,
\label{e1}
\end{equation}
where $c$ is the fragmenting parton (either a gluon or a charm quark).
$D(z,\mu)$ is the fragmentation function and $z$ is
the fraction of the momentum of the parent parton carried by the
charmonium state. Because the gluon or the $c$-quark fragments into
a heavy quarkonium state, the fragmentation function can be computed
perturbatively, in the same spirit as in the colour-singlet model.
This yields the fragmentation function at an initial scale $\mu_0$
which is of the order of $m_c$. If the scale $\mu$ is chosen to be
of the order of $p_T$, then large logarithms in $\mu/m_c$ appear
which have to be resummed using the usual Altarelli-Parisi equation:
\begin{equation}
\mu {\partial \over \partial\mu} D_{i\rightarrow (J/\psi,\chi_i)} 
(z) = \sum_j\int_{z}^1{dy \over y} P_{ij}({z\over y},\mu)
D_{j\rightarrow (J/\psi,\chi_i)}(y) ,
\label{e2}
\end{equation}
where the $P_{ij}$ are the splitting functions of a parton $j$
into a parton $i$. Only the fragmentation of gluons and
charm quarks need be considered since the light quark contributions 
are expected to be very small. The fragmentation functions for
the fragmentation of gluons and charm quarks at the initial scale
$\mu_0$ have been calculated \cite{bryu, bryu2, brcyu, chen, yuan}
and using these as inputs several authors \cite{jpsi} have
computed the contributions coming from fragmentation to the total
$J/\psi$ yield and have found that the gluon fragmentation contribution
significantly increases the cross-section, and the order-of-magnitude
discrepancy between the theory and the data from
the CDF experiment can be resolved. The analyses of the CDF data,
thus, demonstrates the importance of the fragmentation mechanism
as an important source of $J/\psi$ production at large $p_T$.
It is important to study this mechanism in other processes. Some
studies have already been made and fragmentation is predicted to 
have significant effects in $J/\psi$ photoproduction
at large-$p_T$ at HERA \cite{ours1}, and in $J/\psi$ pair production
at the Tevatron \cite{bfp}. A preliminary study \cite{ehv} of higher-order
corrections to the fragmentation contribution in the case of ${}^1S_0$
production has been made, and shows the possible importance of
the higher-order corrections to fragmentation.
 
\vskip10pt
Another important aspect of the physics of quarkonia 
revealed by the analyses of the
CDF data is the importance of colour-octet contributions, which
are ignored in the colour-singlet model because in this model
the relative velocity, $v$, between the heavy quarks in the bound 
state is ignored. However $v$ is, in general, not negligible 
and O$(v)$ corrections need to be taken into account. 
A systematic formulation based on non-relativistic QCD, using the 
factorisation method has been recently carried out \cite{bbl}, and 
in this formulation the quarkonium wave-function admits of a 
systematic expansion in powers of $v$ in terms of Fock-space 
components~: for example, the $\chi$ states have the 
colour-singlet $P$-state component at leading order, but there exist 
additional contributions at non-leading order in $v$, which involve octet 
$S$-state components; i.e. 
\begin{equation}
  \vert \chi_J \rangle = O(1) \vert Q\bar Q \lbrack {}^3P_J^{(1)} \rbrack
      \rangle + O(v) \vert Q\bar Q \lbrack {}^3S_1^{(8)} \rbrack g
      \rangle + \ldots
      \label{e3}
\end{equation}
The importance of the colour-octet components has already been seen in
the analysis of the decays of the $\chi$ states \cite{bbl2}
where the colour-singlet analysis \cite{bgr} reveals a logarithmic infrared
singularity. But the octet component allows the infrared singularity
to be absorbed via a wave-function renormalisation, without having
to introduce an arbitrary infrared cut-off. Thus, a consistent
perturbation theory of $\chi$ decays necessiates the inclusion of
the octet component of the wave-function. As in the case of decays,
the $P$-state fragmentation functions are also infra-red divergent
\cite{bryu2} and, hence, they include the octet component.

\vskip10pt
For $S$-state resonances like the $J/\psi$ and the 
$\psi^{\prime}$, the octet contribution is suppressed by powers of
$v$. Further, the $S$-wave amplitude is not infrared divergent 
and can, therefore, be described in terms of a single colour-singlet
matrix-element. But recently, the CDF collaboration has measured \cite{cdf2}
the ratio of $J/\psi$'s coming from $\chi$ decays to those produced
directly and it turns out that the direct $S$-state production is much
larger than the theoretical estimate. It has been suggested \cite{cgmp}
that a colour octet component in the $S$-wave production coming from
gluon fragmentation as originally proposed in Ref.~\cite{brfl}, can
explain this $J/\psi$ anomaly. 
This corresponds to a virtual gluon fragmenting into an octet ${}^3S_1$
state which then makes a double E1 transition into a singlet ${}^3S_1$
state. While this process is suppressed by a factor of $v^4$ as
compared to the colour-singlet process, it is enhanced by a factor
of $\alpha_s^2$. One can fix the value of the colour-octet matrix-element 
by normalising to the data on direct $J/\psi$ production cross-section 
from the
CDF experiment. The colour-octet contribution to $S$-state production
has also been invoked \cite{brfl} to explain the large $\psi^{\prime}$ 
cross-section measured by CDF \cite{cdf}, but there can be a large
contribution to this cross-section coming from the
decays of radially excited $P$-states \cite{psip}.
Independent tests of the $S$-state colour octet enhancement are
important and recent work shows that a different linear combination
of the same colour octet-matrix elements that appear in the Tevatron
analysis also appears in the analyses of photoproduction \cite{photo}
and hadroproduction experiments \cite{hadro}. These analyses provide
an important cross-check on the colour-octet picture and are very 
important in constraining the magnitude of the octet contributions.

\vskip10pt
In this letter, we examine the implications of these two new
physics aspects of quarkonium production $viz.,$ fragmentation 
and the colour-octet contributions, for $J/\psi$ and $\psi^{\prime}$
production at the LHC. At the LHC energy, we expect gluon
fragmentation to be the most important source of charmonium
production at large $p_T$, and the knowledge of the fragmentation
mechanism and the octet wave-functions gleaned from the study
of charmonium resonances at the Tevatron can be tested at the
LHC. 
The large-$p_T$ production cross-section for the fusion process is
given as
\begin{eqnarray}
&&{d\sigma \over dp_T}(AB \rightarrow (J/\psi,\chi_i) X)
=  \nonumber \\
&& \sum \int dy\int dx_1 x_1G_{a/A}(x_1) x_2G_{b/B}(x_2)
 {4p_T \over 2x_1 -\bar x_T e^y}
{d\hat \sigma \over d \hat t}(ab \rightarrow
(J/\psi,\chi_i) c) .
\label{e4}
\end{eqnarray}
In the above expression, the sum runs over all the partons
contributing to the subprocesses $ab \rightarrow (J/\psi,\chi_i) c$;
$G_{a/A}$ and $G_{b/B}$ are the distributions of the partons $a$ and $b$
in the hadrons $A$ and $B$ with momentum fractions $x_1$ and
$x_2$, respectively. Energy-momentum conservation determines
$x_2$ to be
\begin{equation}
x_2= {x_1 \bar x_T e^{-y} - 2 \tau \over 2x_1-\bar x_T e^y},
\label{e5}
\end{equation}
where $\tau = M^2/s$, with $M$ the mass of the resonance, $s$
the centre-of-mass energy and $y$ the rapidity at which the resonance
is produced.
\begin{equation}
\bar x_T= \sqrt{x_T^2 + 4\tau} \equiv {2M_T \over \sqrt{s}},
\hskip20pt x_T={2p_T \over \sqrt{s}}
\label{e6}
\end{equation}
The expressions for the subprocess cross-sections, $d\hat\sigma/d\hat t$,
are given in Refs.~\cite{br} and \cite{gtw}. 
 
For the fragmentation process, the cross-section is given by a
formula similar to Eq.~\ref{e4} but with an extra integration
over $z$, or equivalently over $x_2$. We have
\begin{eqnarray}
&&{d\sigma \over dp_T}(AB \rightarrow (J/\psi,\chi_i) X)
=  \nonumber \\
&& \sum \int dy dx_1 dx_2 G_{a/A}(x_1) G_{b/B}(x_2)
D_{c\rightarrow (J/\psi,\chi_i)} (z)
{2p_T \over z} {d\hat \sigma \over d \hat t}(ab \rightarrow cd) ,
\label{e7}
\end{eqnarray}
with $z$ given by
\begin{equation}
z= {\bar x_T \over 2} ({e^{-y} \over x_2} + {e^y \over x_1}) .
\end{equation}
For $d\hat\sigma/d\hat t(ab \rightarrow cd)$, we have used the
lowest-order expressions.
For the fragmentation functions at the initial scale $\mu=\mu_0$, we
use the results of Refs.~\cite{bryu} and \cite{bryu2} for the
gluon fragmentation functions
into $J/\psi$ or $\chi$ states, and Refs.~\cite{brcyu}, \cite{chen} and 
\cite{yuan} for the corresponding fragmentation functions
of the charm quark. These
fragmentation functions include 
the colour-octet component in the $P$-state, but do not
include any colour-octet contribution in the $S$-state. For the
case of gluon fragmentation, we have included the effect
of the $S$-state colour-octet component by modifying the fragmentation
functions as in Ref.~\cite{brfl}. For the charm fragmentation,
the $S$-state colour-octet contributions are sub-dominant and we have
neglected these contributions.

\vskip10pt
We have computed the cross-sections for the planned LHC energy
$\sqrt{s}=14$~TeV using the MRSD-${}^\prime$ parton densities \cite{mrs} 
In Fig.~1, we present the $J/\psi$ production cross-section 
$Bd\sigma / dp_T$ as a function of $p_T$, where $B$
is the branching ratio of the $J/\psi$ into leptons. We have assumed
a rapidity coverage $-2.5 \le y \le 2.5$ and integrated over the
full rapidity interval. The parton densities are evolved to a 
scale $Q = \mu/2$, where $\mu$ is chosen to be $M_T$ for the
fusion case and equal to $p_T^{g,c} = p_T/z$ for the fragmentation
case. The fragmentation functions are evolved to the scale $p_T/z$.
We find that the cross-section for $J/\psi$ production is completely
dominated by the gluon fragmentation contribution (shown by the
dotted line in Fig.~1) and is larger than the fusion contribution
and the charm-quark fragmentation contribution (shown by the solid
and the dashed lines respectively) by about two orders of magnitude.
The cross-section for $J/\psi$ production is large and even at
$p_T=100$~GeV, the cross-section is as large as 0.1pb. Assuming 
a value of luminosity $\sim 10^4$~pb${}^{-1}$, typical of that 
expected at the LHC, we would expect of the order of $10^3$ events
even for a $p_T$ of 100~GeV. For values of $p_T$ so much larger
than the charm quark mass, the fragmentation picture becomes 
exact and the prediction of the cross-section using the fragmentation
picture should be equal to the experimentally measured cross-section. 
The experimental measurement of the $J/\psi$ cross-section will,
therefore, be a crucial test of the fragmentation picture. Further,
for these predictions we have used precisely the same values of
the non-perturbative inputs i.e. the colour-octet matrix elements
that were determined from the analysis of the CDF data. 
Therefore, the LHC measurement will also be a test of the magnitude
of the colour-octet contributions.

In Fig.~2, we have shown the cross-section for $\psi^{\prime}$
production at LHC energies. Again, the cross-section is dominated
by the octet-enhanced gluon fragmentation contribution. The overall
magnitude of the cross-section is smaller than in the case of $J/\psi$
production, but one still expects a comfortably large event rate
for $\psi^{\prime}$ production for $p_T \sim 100$~GeV.

In conclusion, we have studied the contribution to large-$p_T$
$J/\psi$ and $\psi^{\prime}$ production at the LHC energy, coming
from fragmentation of gluons and charm quarks. We find that the
cross-section is overwhelmingly dominated by the gluon fragmentation
contribution. The full cross-section is large and suggests a significant
event rate even for values of $p_T$ much larger than 100~GeV. The
measurement of the charmonium cross-sections at LHC will provide
a crucial test of the fragmentation picture and provide a check on the
magnitude of the colour-octet matrix elements.

\newpage
\begin{figure}[htb]
\vskip 8in\relax\noindent\hskip -1in\relax{\includegraphics{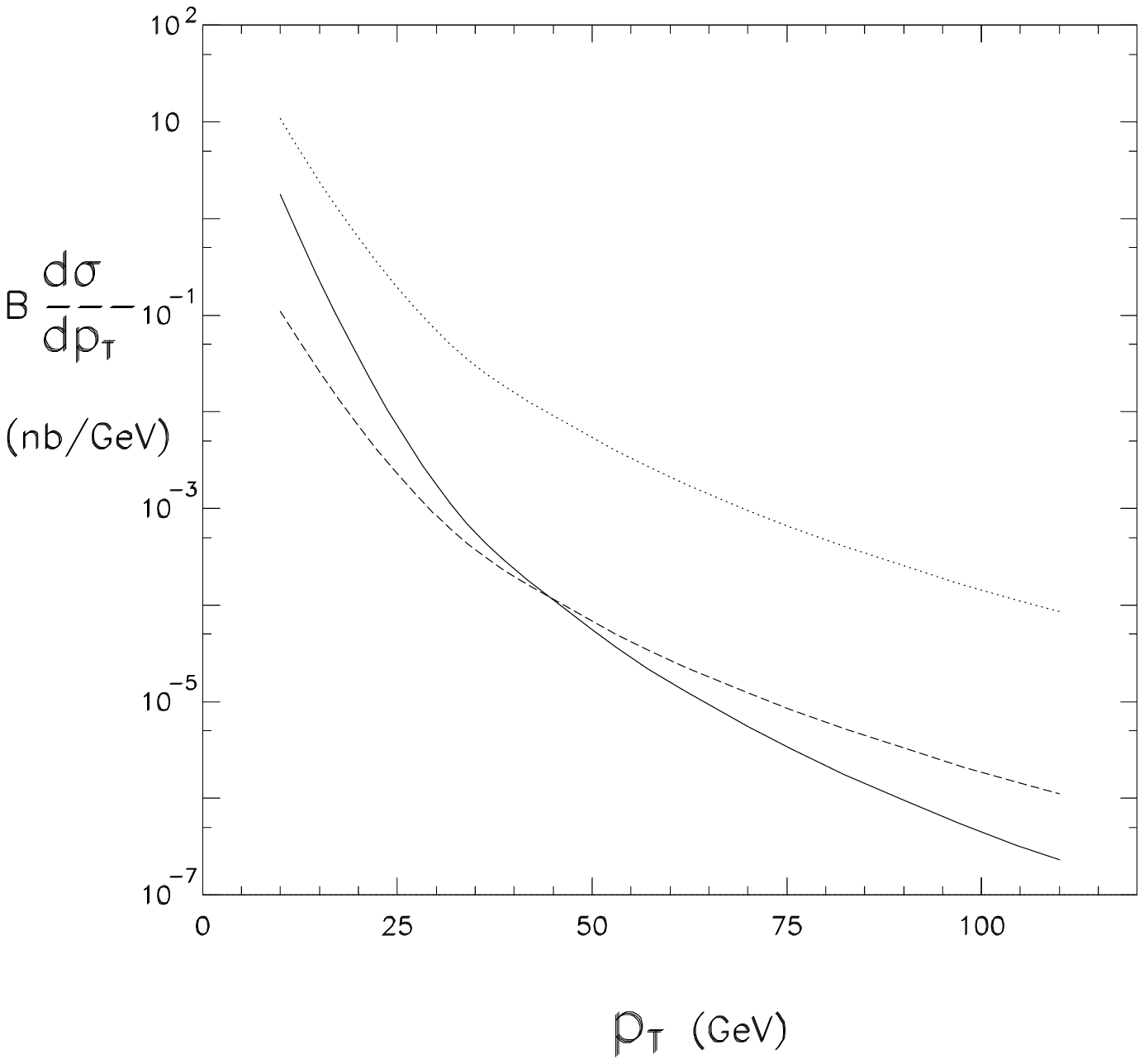}}

\vspace{-20ex}
\caption{$Bd\sigma/dp_T$ (in nb/GeV) for $J/\psi$ production at
14~TeV c.m. energy with $-2.5 \le y \le 2.5$. The
solid, dashed and dotted lines represent the fusion, charm quark
fragmentation and gluon fragmentation contributions.}
\end{figure}
\begin{figure}[htb]
\vskip 8in\relax\noindent\hskip -1in\relax{\includegraphics{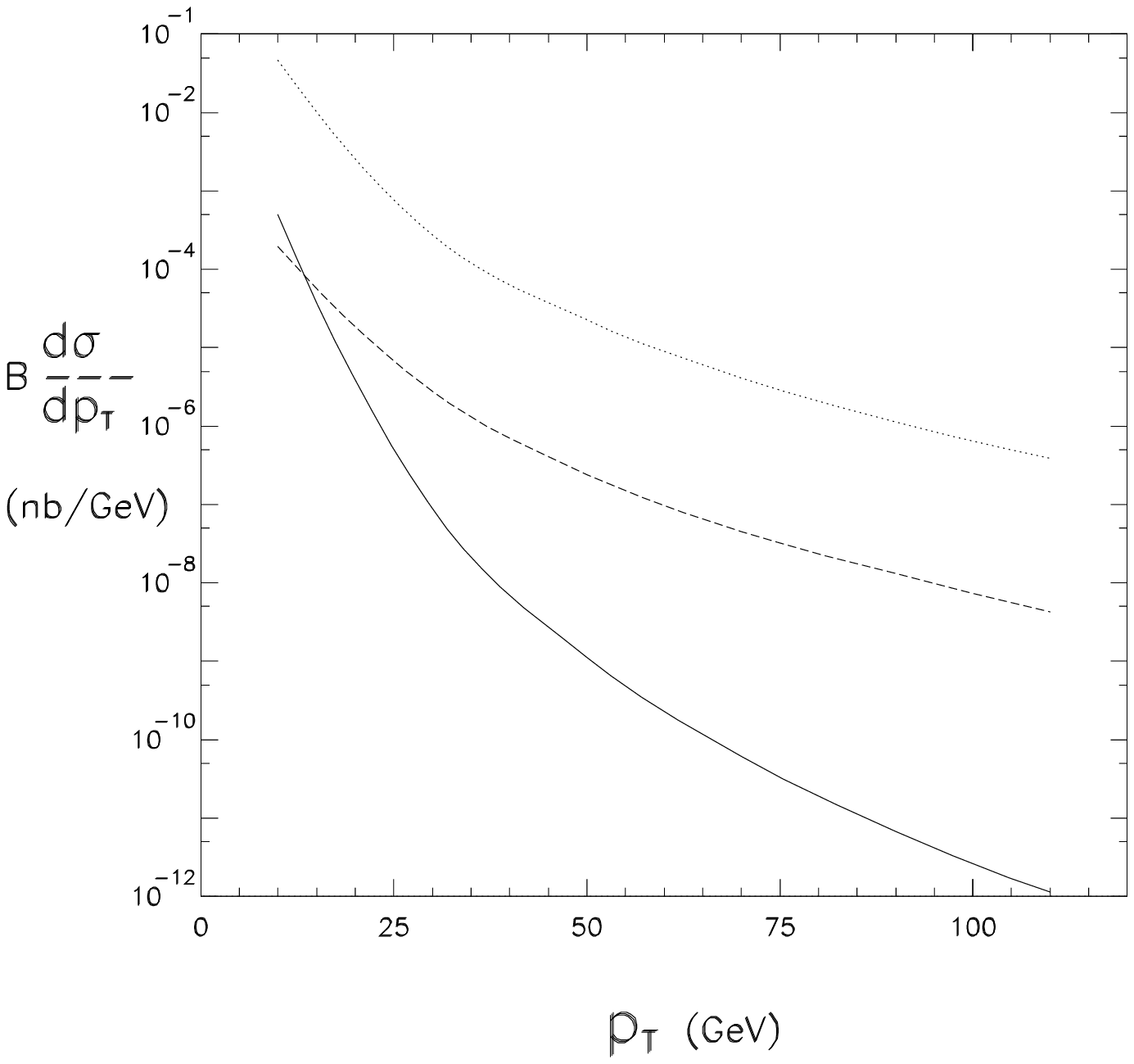}}

\vspace{-20ex}
\caption{$Bd\sigma/dp_T$ (in nb/GeV) for $\psi^{\prime}$ production
at 14~TeV c.m. energy with $-2.5 \le y \le 2.5$. The
solid, dashed and dotted lines represent the fusion, charm quark
fragmentation and gluon fragmentation contributions.}
\end{figure}
\end{document}